\documentclass[aps,prl,twocolumn,superscriptaddress]{revtex4}
\usepackage{amsmath,amssymb,graphicx,color,braket,hyphenat}
\usepackage[normalem]{ulem}
\newcommand{\cdit}{\!\cdot\!}
\newcommand{\minus}{\!-\!}

\newcommand{\Mod}[1]{\left|#1\right|}
\def\half{\frac{1}{2}}
\def\phi{\varphi}

\begin{document}

\title{Multiple observers can share the non-locality of half of an entangled pair \\ by using optimal weak measurements}

\author{Ralph Silva}\affiliation{H.H. Wills Physics Laboratory, University of Bristol, Tyndall Avenue, Bristol, BS8 1TL, U.K.}
\author{Nicolas Gisin}\affiliation{Group of Applied Physics, University of Geneva, CH-1211 Geneva 4, Switzerland}
\author{Yelena Guryanova}\affiliation{H.H. Wills Physics Laboratory, University of Bristol, Tyndall Avenue, Bristol, BS8 1TL, U.K.}
\author{Sandu Popescu}\affiliation{H.H. Wills Physics Laboratory, University of Bristol, Tyndall Avenue, Bristol, BS8 1TL, U.K.}

\begin{abstract}
We investigate the trade-off between information gain and disturbance for von Neumann measurements on spin-$\half$ particles, and derive the measurement pointer state that saturates this trade-off, which turns out to be highly unusual. We apply this result to the question of whether the non-locality of a single particle from an entangled pair can be shared among multiple observers that act sequentially and independently of each other, and show that an arbitrarily long sequence of such observers can all violate the Clauser-Horne-Shimony-Holt$\hyp$Bell inequality.
\end{abstract}

\maketitle

\emph{Introduction.---} A central paradigm in quantum theory is that measurements are necessarily disturbing $\hyp$ in order to probe the properties of a system one must perturb it\cite{Heisenberg}. The measurement postulate \cite{vNpost} states that performing what is referred to as a ``strong" measurement collapses the system into one of the eigenstates of the measured observable; this type of measurement offers the maximum information about the system.

On the other hand, there exist measurement schemes that disturb the system infinitesimally, offering only a small amount of information about the state. Such ``weak" measurements are often considered in conjunction with post-selection\cite{ABL}, a formalism that has precipitated the study of weak values\cite{weakvalue}. Of course, these measurements are important by themselves, even without post-selection. Indeed all macroscopic measurements are weak measurements\cite{macro,magnet}.

Here we consider measurements of all intermediate strengths focusing on the trade-off between the degree of disturbance and the amount of information we gain about the system. This trade-off has been explored extensively, both in the context of specific measuring devices\cite{prior,qubitpointer} and abstract measurement representations\cite{Fuchs,Banaszek,Horodecki,recent}.

The subject of our investigation is the von Neumann type measurement, which is characterized by the pointer of the measuring device being displaced proportionally to the value of the measured observable. This offers arguably the most direct connection between the measured physical quantity and the reading of the measuring device. We are interested in deriving the optimal measurements, i.e. those that maximize the information gain for a given disturbance to the system.

In this Letter, we consider the case of dichotomic measurements on spin-$\half$ particles.

The information gain and disturbance can be modified by changing the initial state of the pointer as well as the strength of the coupling between the system and the measuring device. However, the optimal information gain vs. disturbance trade-off cannot be achieved by only tuning the coupling strength (which is equivalent to re-scaling the state of the pointer). Rather the initial state of the pointer must be appropriately chosen. We determine the optimal pointer state, and find it to be very counter-intuitive. In particular, it is nothing like the Gaussian wave packet that is almost universally considered and considerably outperforms it.

We then use a simple bipartite scenario involving successive measurements, to find a constraint on the trade-off, in a similar vein to those derived in \cite{Banaszek,recent}. The trade-off attained by the optimal pointer saturates this constraint.

Since von Neumann measurements are, on the one hand rich enough to allow us to tune this trade-off, and on the other hand simple enough to allow manageable calculations, they enable us to raise and answer a new fundamental question in non-locality: can the non-locality of an entangled pair of particles be distributed among multiple observers, that act sequentially and independently of each other? We consider the scenario that a single observer has access to one of the particles of an entangled pair, and a group of observers have access to the second particle. Each observer in the second group acts independently, performing a measurement on the particle before passing it on to the next member of the group. We address the question of whether the single observer with the first particle can see non-local correlations with \emph{all} of the members in the second group.

Crucially, we find that each member in the second group cannot perform a very weak measurement, since this is unable to extract enough information to observe non-local correlations. Hence the state is disturbed significantly, and it is not clear that subsequent observers can still observe non-local correlations. Nevertheless, we show that an arbitrary number of independent observers can indeed see consecutive violations of the CHSH (Clauser-Horne-Shimony-Holt)$\hyp$Bell inequality. As well as teaching us about the nature of non-locality, this problem illuminates the nature of the information gain vs disturbance trade-off.

\onecolumngrid
\newpage
\twocolumngrid
\emph{von Neumann measurement pointers for spin-$\half$ particles.---} In a von Neumann type measurement, the pointer is shifted proportional to the eigenvalues of the measured observable
\begin{equation}\label{vonNeumann}
	\ket{\Psi} \otimes \ket{\phi(q)} \longrightarrow \sum_a \braket{a|\Psi} \cdot \ket{a} \otimes \ket{\phi(q-g_0a)},
\end{equation}
where $\Psi$ and $\phi(q)$ are the initial states of the system and pointer, respectively, the index $a$ refers to the eigenbasis of the observable, $q$ is the position of the pointer, and $g_0$ is a coupling constant. The outcome of the measurement is then provided by reading the position of the pointer.

The evolution (\ref{vonNeumann}) is generated by the interaction Hamiltonian $H(t) = g(t) \cdit A\otimes p$, where A is the measured observable, $p$ is the momentum operator of the pointer conjugate to $q$ and $g(t)$ is non-zero only during a short time interval and normalized s.t. $\int g(t)dt = g_0$. Here, we take $g_0=1$, which can be done without loss of generality by simply rescaling the pointer state\cite{footnote1}.

In a strong measurement the pointer's initial state is narrower than the distance between the eigenvalues, i.e. $\braket{\phi(q-a)|\phi(q-a^\prime)}=\delta_{aa^\prime}$, hence reading the pointer's position provides full information of the measured physical quantity and collapses the system into the corresponding eigenstate of the observable.

Conversely, if the pointer spread is very large, covering the entire spectrum of eigenvalues, reading the pointer position provides essentially no information since $\braket{\phi(q-a)|\phi(q-a^\prime)}\approx 1$ and the system is not perturbed,
\begin{align}
	\ket{\Psi^\prime}_{|q_0} &= \sum_a \braket{a|\Psi} \braket{q_0|\phi(q-a)} \ket{a} \nonumber\\
	&\approx \braket{q_0|\phi(q)} \sum_a \braket{a|\Psi} \ket{a} = \braket{q_0|\phi(q)} \ket{\Psi}.
\end{align}
This is the limit of a weak measurement.

We now consider measurements in between the two extremes. Focusing on spin-$\half$ particles, the initial state of the spin in the eigenbasis of the measured observable is $\ket{\Psi} = \alpha\ket{\uparrow} + \beta\ket{\downarrow}$, hence
\begin{equation}
	\ket{\Psi} \otimes \ket{\phi(q)} \longrightarrow \alpha\ket{\uparrow}\otimes\ket{\phi(q-1)} + \beta\ket{\downarrow}\otimes\ket{\phi(q+1)}.
\end{equation}

For simplicity, we consider pointer states with symmetric modulus, i.e. $\Mod{\phi(q)}=\Mod{\phi(-q)}$. We also take $\phi(q)$ to be real-valued, without loss of generality, since complex pointers are shown not to outperform real ones (see Appendix B).

To determine the disturbance produced by the measurement, we compute the system post-measurement state by tracing out the pointer (Appendix A)
\begin{equation}\label{postmeas}
	\rho^\prime = F\ket{\Psi}\!\!\bra{\Psi} + (1\minus F) \left( \pi^+\! \ket{\Psi}\!\!\bra{\Psi} \pi^+\! + \pi^-\!\ket{\Psi}\!\!\bra{\Psi} \pi^- \right),
\end{equation}
where $\pi^+=\ket{\uparrow}\!\!\bra{\uparrow}$ and $\pi^- = \ket{\downarrow}\!\!\bra{\downarrow}$. The quantity $F$ is independent of the state of the spin, and is the scalar product of the displaced pointer states,
\begin{equation}\label{F}
	F=\int_{-\infty}^{+\infty} \!\!\!\!\! \phi(q+1)\phi(q-1) \;dq.
\end{equation}

We call $F$ the `quality factor' of the measurement since it is the proportion of the post-measurement state that corresponds to the original state. The remainder corresponds to the state decohered in the measurement eigenbasis, as it would have been if measured strongly.

The other quantity of interest is the information gain. Since we are measuring a dichotomic observable, we digitize the reading of the pointer, associating positive positions to the outcome $+1$ and negative positions to $-1$ (see discussion in Appendix C). The probability of the outcomes $\pm 1$ is then (Appendix A)
\begin{equation}\label{prob}
	P(\pm 1) = G \braket{\Psi|\pi^\pm|\Psi} + (1-G)\frac{1}{2}.
\end{equation}

$G$ is also independent of the state of the spin, and depends on the width of the pointer compared to the distance between the eigenvalues,
\begin{equation}\label{G}
	G = \int_{-1}^{+1} \!\!\! \phi^2(q) dq.
\end{equation}

The first term in (\ref{prob}) represents the contribution of the probability as if there was a strong measurement, so we call $G$ the precision of the measurement. The other term, $(1-G)\half$, corresponds to a random outcome.

Consider for example the simple case of a square pointer state: $\phi(q) = 1/(\sqrt{2\Delta})$ for $-\Delta<q<+\Delta$ and zero elsewhere.  If the spread $\Delta$ is smaller than 1, then reading the pointer's position provides full information of the measured spin, i.e. $\Delta\!\leq\! 1$ corresponds to a strong measurement: $F\!=\!0$, $G\!=\!1$. When $\Delta\!>\!1$, we find that $G\!=\!1\minus F$. Hence square pointers correspond to measuring strongly with probability $G$ and producing a random result, without measuring, with probability $1\minus G$.

\begin{figure}[h]
\includegraphics[width=0.9\linewidth]{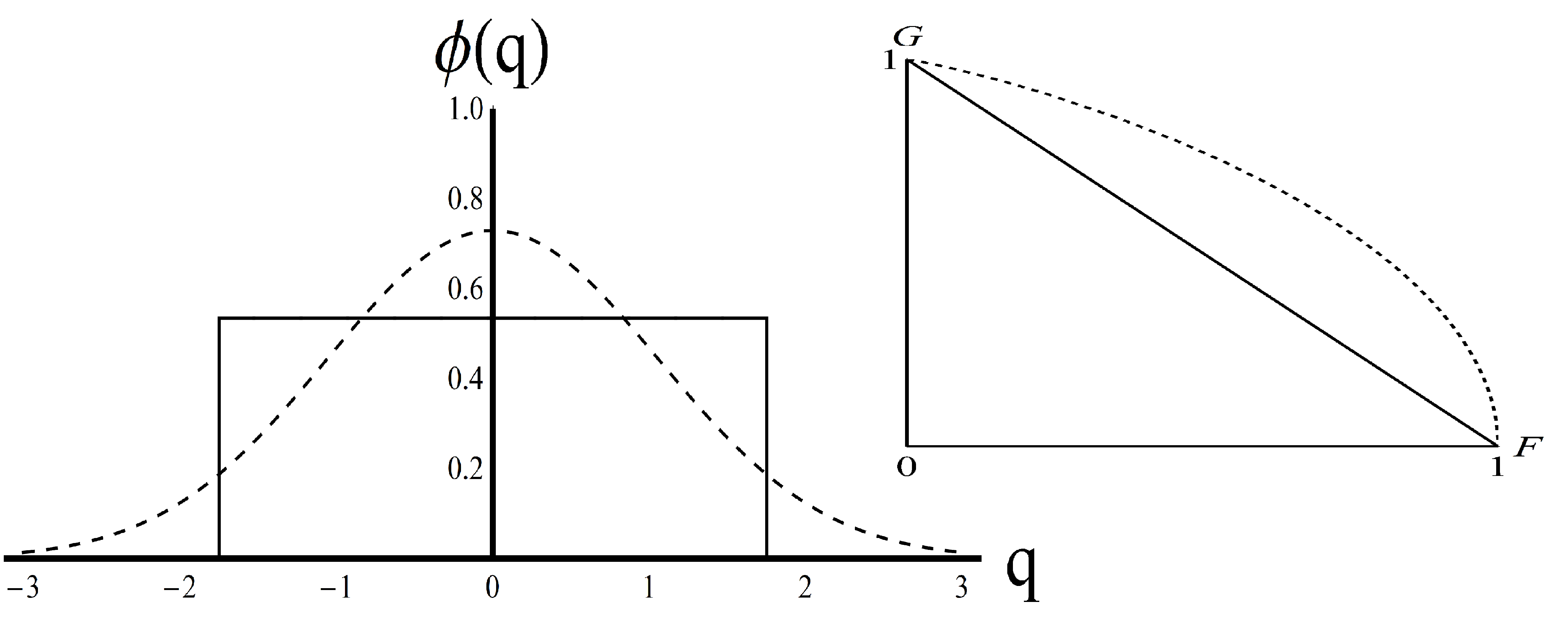}
\vspace{-10pt}
\caption{\label{SqGs}Square (solid line) and Gaussian (dashed line) pointers of equal width $\Delta=1.5$ with (inset) the corresponding trade-off between the precision $G$ and quality factor $F$.}
\end{figure}

However, square pointers are far from optimal: Gaussian wave-packets achieve a better trade-off between $F$ and $G$ (see Fig. \ref{SqGs}), but are still not optimal.

\onecolumngrid
\newpage
\twocolumngrid
\emph{Optimal pointers.---} Since $F$ and $G$ are solely functionals of the pointer state it is natural to look for the one that achieves the best trade-off by using variational calculus (Appendix B).

Interestingly, we find that for any quality factor F, there is an entire family of optimal pointer states that achieve the maximum precision G. Each element of this family is defined by the choice of an arbitrary function $f(q)$ in the interval $-1\!<\!q\!<\!+1$ such that the norm of the pointer state within this interval is the precision G (\ref{G}). The function is then copied to all other regions between adjacent odd points $q\!=\!2n\!-\!1$ and $q\!=\!2n\!+\!1$ with the relative height of the function in each region falling under an exponential envelope that depends on G,
\begin{align}\label{optpointer}
	\phi(q) &= f(q-2n) \left(\sqrt{\frac{1-G}{1+G}}\right)^{|n|} \nonumber\\
	\forall q &\in(2n\!-\!1,2n\!+\!1],\;\; n\in\mathbb{Z}.
\end{align}

\begin{figure}[h]
\includegraphics[width=0.9\linewidth]{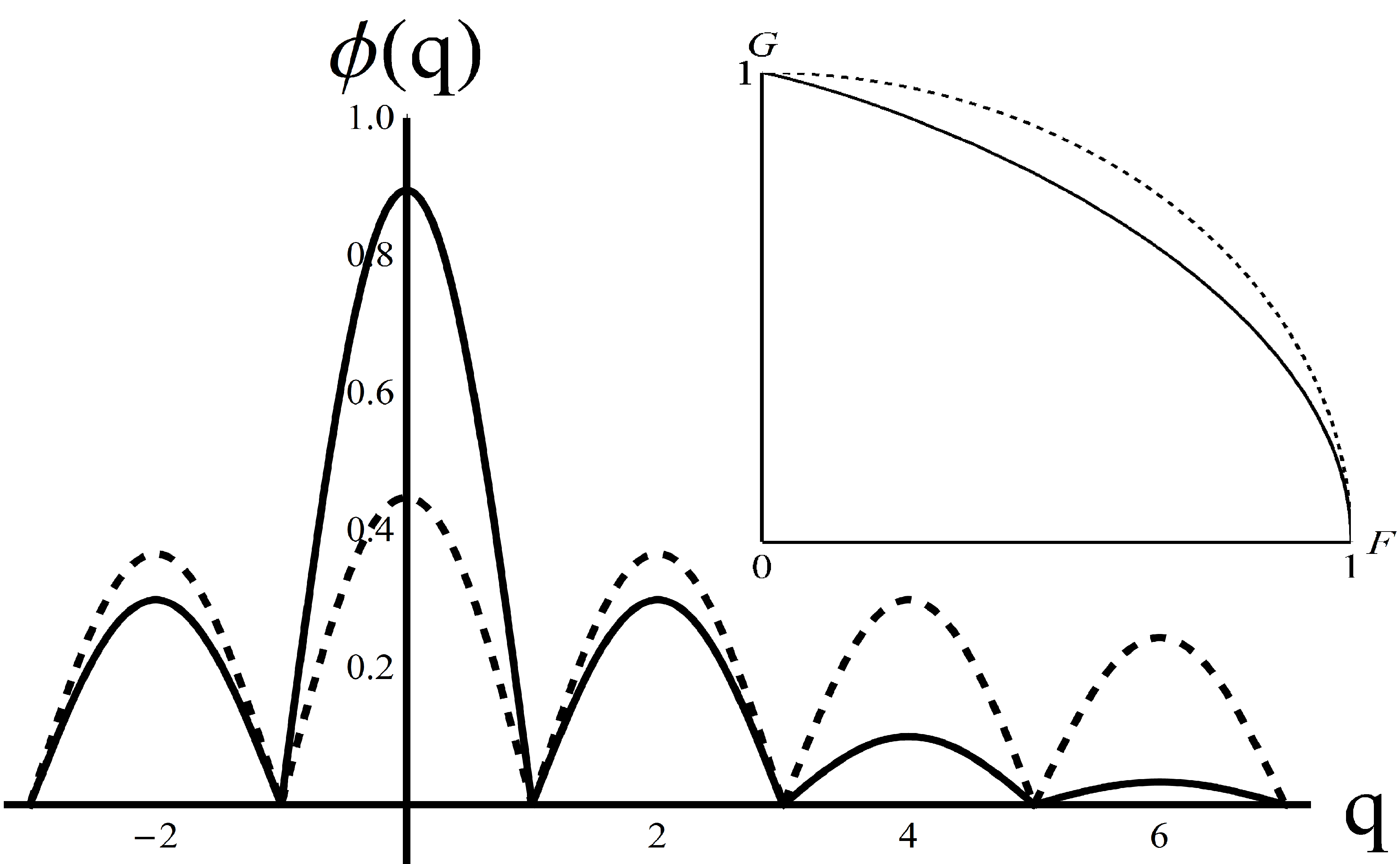}
\caption{\label{optimal}Plot of two optimal pointer distributions, $\{G=0.8,F=0.6\}$ (solid) and $\{G=0.2,F=0.98\}$ (dashed). Inset: Comparison of the optimal trade-off (dashed) to that attained by the Gaussian pointer (solid).}
\end{figure}

Two such optimal pointer states are plotted in Fig. \ref{optimal}, along with the trade-off compared to that of Gaussian pointers. For an optimal pointer state, the trade-off is given by
\begin{equation}\label{tradeoff}
	F^2 + G^2 = 1.
\end{equation}

\begin{figure}[h]
\includegraphics[width=0.8\linewidth]{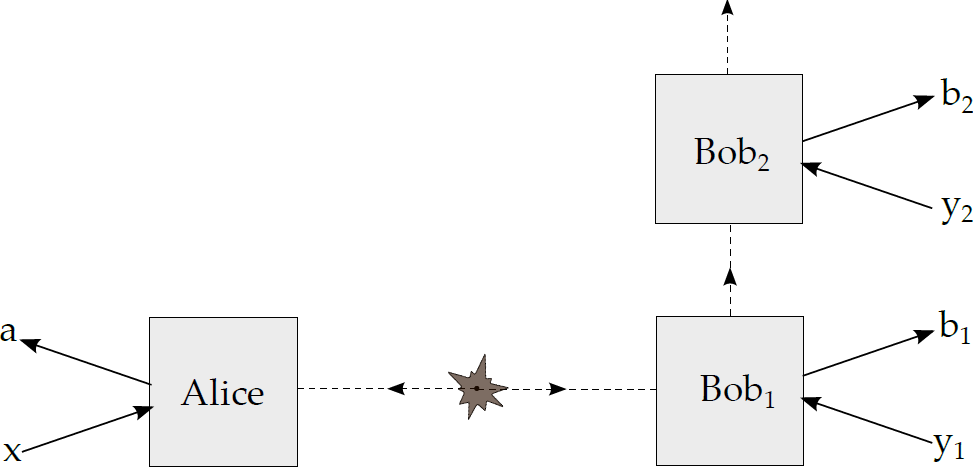}
\caption{\label{Bellscenario}Bell scenario involving a single Alice and multiple Bobs, where the dashed lines indicates a spin-$\half$ particle being transmitted, and the solid lines the inputs and outputs.}
\end{figure}

\bigskip
\emph{A bound on the disturbance-precision trade-off.---} Interestingly, the above trade-off (\ref{tradeoff}) can also be deduced from a simple Bell inequality type scenario (Fig. \ref{Bellscenario}).  Alice and Bob each possess one half of a singlet state of spin-$\half$ particles. Alice receives a binary input $x\in\{0,1\}$, and performs a strong projective measurement of her spin along a corresponding direction $\bar{u}_x$; we label her outcome $a=\pm 1$. Bob receives two consecutive binary inputs $y_1,y_2\in\{0,1\}$, and performs two consecutive spin measurements along corresponding directions $\bar{w}_{y_1}$ and $\bar{v}_{y_2}$; his outputs are labelled $b_1$ and $b_2$ ($\pm 1$). Bob's first measurement has intermediate strength, while his second is a strong measurement.

Such a scenario is characterized by the conditional probabilities of the outcomes, $P(ab_1b_2|xy_1y_2)$. To calculate these, we require the state of Bob's spin after his first measurement of intermediate strength. This is different from the state in Eq. (\ref{postmeas}), since here we require the post-measurement state \emph{given the specific outcome} $b_1$, and thus trace only over either positive or negative pointer positions respectively.

We find that the outcome dependent state of the spin-$\half$ particle is dependent on both the quality factor $F$ and precision $G$ of the measurement
\begin{equation}
	\rho^\prime_{|b_1} \!=\! \frac{F}{2}\rho + \left(\!\frac{1+b_1G\minus F}{2}\!\right)\pi^+\!\rho\pi^+ + \left(\!\frac{1\minus b_1G\minus F}{2}\!\right)\pi^-\!\rho\pi^-,
\end{equation}
where $\pi^+$ and $\pi^-$ denote the projectors of the spin measurement, $\rho$ is the pre-measurement state and $\rho^\prime_{|b_1}$ is the unnormalized post-measurement state of the system given the outcome $b_1$. From this state, one arrives at the conditional probability (Appendix D),
\begin{align}\label{condprob}
&P(ab_1b_2|xy_1y_2) = \frac{b_1G}{4} \left(\! \frac{a\; \bar{u}_x\!\cdit\bar{w}_{y_1} + b_2\; \bar{w}_{y_1}\!\cdit\bar{v}_{y_2}}{2} \!\right) + \nonumber \\
&\frac{F}{4} \left(\! \frac{1+ab_2\;\bar{u}_x\cdit\bar{v}_{y_2}}{2} \!\right) \!+\! \left(\!\frac{1\minus F}{4}\!\right) \!\left(\! \frac{1+ab_2\; \bar{u}_x\!\cdit\bar{w}_{y_1} \bar{w}_{y_1}\!\cdit\bar{v}_{y_2}}{2} \!\right),
\end{align}
which is non-signalling between Alice and Bob, as expected.

Furthermore, being a probability, it must lie between $0$ and $1$. Choosing the measurement directions to be $\bar{u}_0 = \bar{Z}$, $\bar{w}_0 = -\bar{X}$, and $\bar{v}_0 = \bar{Z}\sin\theta - \bar{X}\cos\theta$ (where $\bar{Z}$ and $\bar{X}$ are two orthogonal directions in space), along with the outcomes $a=b_1=b_2=1$, we obtain the inequality	$P(111|000) = F\sin\theta + G\cos\theta \leq 1$. This is the expression of a tangent to the unit circle $F^2+G^2=1$, at the point $\{\sin\theta,\cos\theta\}$. Varying over $\theta$, we obtain all of the tangents to the unit circle as constraints on the pair $\{F,G\}$, and thus the pair must lie within the unit circle. The optimal pointer described previously saturates this constraint.

Using such a Bell scenario to examine the trade-off is a natural method to study weak measurements in generalized probability theories, where one can expect the optimal trade-off to differ from the quantum trade-off.

\bigskip
\emph{Consecutive violations of the CHSH-Bell inequality.---} Armed with an understanding of the trade-off between information gain and disturbance, we now raise a novel and fundamental question in non-locality $\hyp$ can multiple observers share the non-locality present in a single particle from an entangled pair? To answer this question, we consider the Bell scenario in Fig. \ref{Bellscenario}, where Alice has one half of an entangled pair of spin-$\half$ particles, but instead of a single Bob performing two consecutive measurements, there are two Bobs that each perform a measurement one after the other on the second particle of the pair. The Bobs are independent, i.e. Bob$_2$ is ignorant of the direction that Bob$_1$ measures his spin in as well as the outcome of his measurement.

We investigate whether the statistics of the measurements of Bob$_1$ and Bob$_2$ can both be non-local with Alice by testing the conditional probabilities $P(ab_1|xy_1)$ and $P(ab_2|xy_2)$ against the CHSH inequality\cite{CHSH}.

At first one may think it impossible to have simultaneous violations Alice-Bob$_1$ and Alice-Bob$_2$ because of the monogamy of entanglement\cite{monogamy} and of non-locality \cite{MAG,nonlocal}. However, these results assume no-signalling between all parties, while in our scenario Bob$_1$ implicitly signals to Bob$_2$ by his choice of measurement on the state before he passes it on. Hence, no monogamy argument holds, and one has to look more closely at the situation.

An unusual feature of this Bell scenario is that Bob$_2$'s CHSH value depends on the \emph{input bias} of Bob$_1$, i.e. the frequency with which Bob$_1$ received the input $0$ versus the input $1$. Even though the CHSH expression contains only conditional probabilities, the state that Bob$_2$ measures has been perturbed by Bob$_1$. Since Bob$_2$ is independent of Bob$_1$, his density matrix is the mixture of the states given each of Bob$_1$'s two possible measurements, weighted by their relative frequencies. Hence the input bias of Bob$_1$ affects the statistics of Bob$_2$'s measurement.

To begin with, we assume the measurements are unbiased, i.e. both Bob's receive the inputs $0$ and $1$ with equal probability. Clearly Bob$_1$ cannot perform a strong measurement, since he would destroy the entanglement, and prevent Bob$_2$ from being non-local with Alice. However, Bob$_1$ may not be able to observe non-locality with a very weak measurement either. To see this precisely, consider that Alice and the Bobs initially share a singlet state, and that they perform the standard measurements that attain Tsirelson's bound for the CHSH inequality: i.e. Alice measures in the $\bar{Z}$ or $\bar{X}$ direction, corresponding to inputs $0$ or $1$ respectively, and the Bobs measure in the directions $-(\bar{Z}+\bar{X})/\sqrt{2}$ or $(-\bar{Z}+\bar{X})/\sqrt{2}$, for their respective inputs $0$ or $1$.

\begin{figure}[h]
\includegraphics[width=1.0\linewidth]{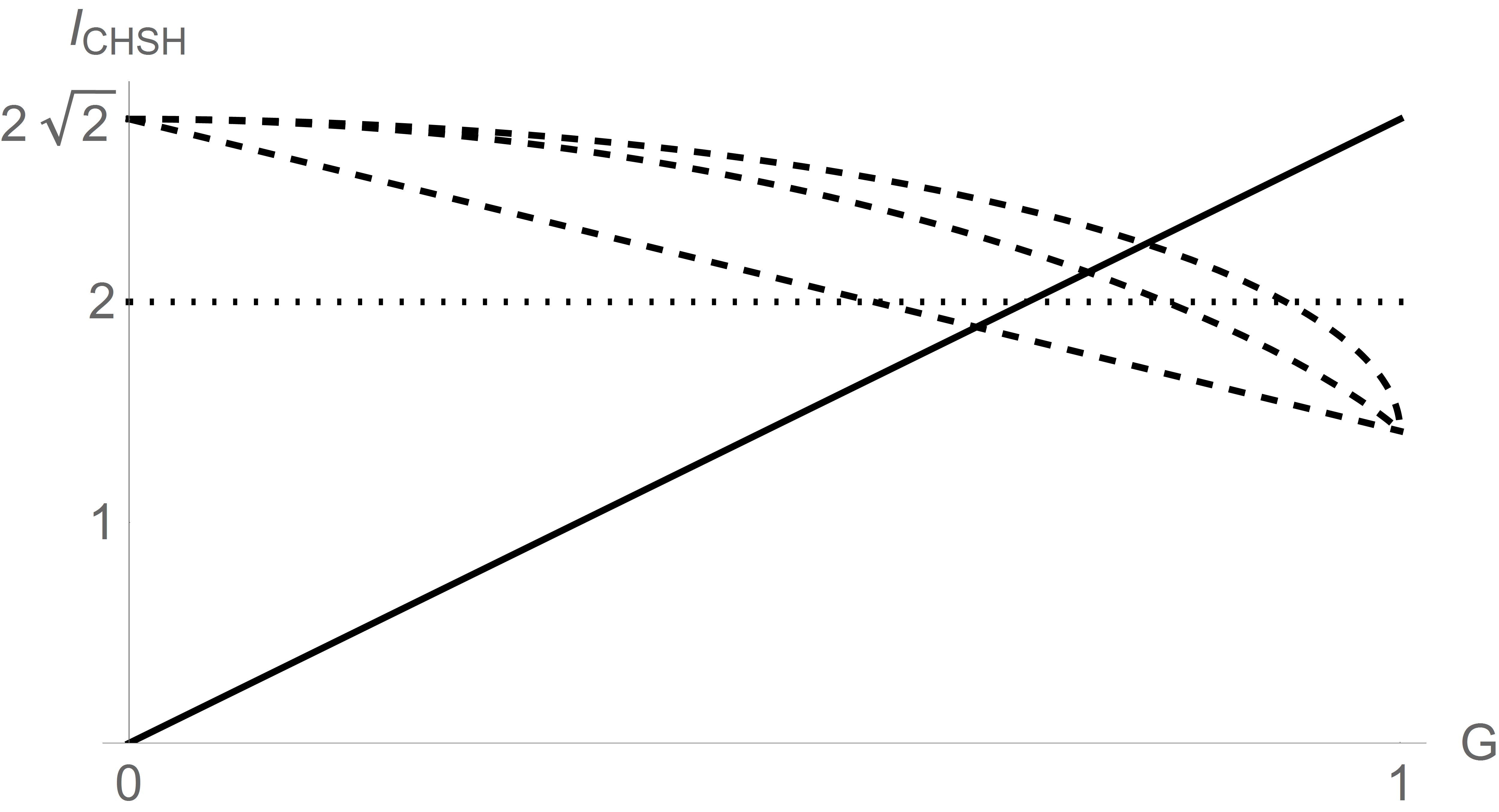}
\caption{\label{diffpointerBell} Plot of $I^{(1)}_{CHSH}$ (solid) as a function of the precision $G$ of Bob$_1$, together with $I^{(2)}_{CHSH}$ (dashed) for different pointer types, (from bottom) square, Gaussian and optimal.}
\end{figure}

Using the form of the CHSH expression\cite{CHSH} with the classical bound at $2$ and the quantum bound at $2\sqrt{2}$, we find that the CHSH values of Alice with each Bob are given by $I^{(1)}_{CHSH} = 2\sqrt{2}G$, and $I^{(2)}_{CHSH} = \sqrt{2}(1+F)$, where $G$ and $F$ are the precision and quality factor of Bob$_1$'s measurement. These are plotted in comparison to the classical bound in Fig. \ref{diffpointerBell}.

We see from the figure that Bob$_1$ must tune the precision of his measurement, as either a strong or weak measurement would prevent Bob$_2$ or himself, respectively, from seeing a CHSH violation. He must also use a pointer with a good trade-off $\hyp$ one cannot have a double violation  using a square pointer, while it is possible with a Gaussian or optimal pointer.

\bigskip
\emph{Longer sequences of CHSH violations with biased inputs.---} Since it is possible to have two Bobs simultaneously violate CHSH with Alice, the next natural question is whether there is a limit to the number of consecutive violations achievable.

We find that it is possible for more than two Bobs to violate CHSH with Alice, if the frequency of the inputs $0$ and $1$ to each Bob is not the same\cite{footnote2}. In Appendix F, we provide an explicit measurement protocol that does so in the case that one of the inputs to the various Bobs occurs much more often than the other input. In this scenario, there is no limit to the number of Bobs that can violate CHSH with Alice $\hyp$ the larger the bias of the inputs, the longer the sequence of violations. However, in our protocol, the value of the CHSH violation in the sequence falls off super-exponentially: if $V_n = I^{(n)}_{CHSH}-2$ is the maximum violation that can be achieved by Bob$_n$ with Alice, we find that for large $n$, (Appendix G)
\begin{equation}
	V_{n+1} \approx \frac{V_n^3}{4}
\end{equation}

\onecolumngrid
\newpage
\twocolumngrid
\emph{Discussion and open problems.---} We have seen that the trade-off between information gain and disturbance for von Neumann measurements is strongly dependent on the initial state of the pointer, and the optimal pointer state differs considerably from the pointers considered usually, such as the Gaussian wave-packet. An interesting question to ask is what form the optimal pointer takes for measurements on higher dimensional systems.

We also obtained a constraint on the trade-off by relating it to the probabilities in a simple Bell scenario. Such a method can be used to extend the concept of weak measurements to general non-local theories.

In the case of multiple observers violating a Bell inequality, we have numerical evidence that if the inputs to the various Bobs are unbiased, it is impossible to have more than a double violation of CHSH with Alice. Proving this analytically is an open problem. For general input bias, an open question is whether there exists a protocol that achieves a better CHSH violation than that found in this Letter. Also, one may generalize to the case when the Bobs have some information about each others' inputs and/or outcomes, this will presumably improve the CHSH violation. Finally, it would be interesting to include multiple Alices in the setup, and investigate if it is possible to have an arbitrarily long sequence of pairs of Alices and Bobs that violate a Bell inequality.

\bigskip
\emph{Acknowledgements.---} We thank Anthony J. Short for stimulating discussions during the course of this work. We also acknowledge financial support by the European projects ERC-AD MEC, ERC-AD NLST and Chist.Era DIQIP, the Swiss project NCCR-QSIT, and the EPSRC.

\onecolumngrid
\newpage

\section{Appendix}

\section{A. Quantifying the information gain and disturbance of a weak von Neumann measurement on a spin-$\half$ particle.}\label{AppA}

For this measurement, a continuous pointer serves as the measuring device, and the outcome of the measurement is read from the position of the pointer, denoted by $q$. The spin observable being measured (denoted by the operator $\sigma$) is coupled to the momentum $p$ of the pointer, via the interaction Hamiltonian $H_{int}(t) = g(t) \sigma \otimes p$. With a suitable choice of the impulse $g(t)$, the evolution of the system and pointer state is described by the unitary
\begin{equation}\label{unitary}
	\hat{U} = e^{i \sigma \otimes p}.
\end{equation}

This unitary is a displacement operator on the pointer, that shifts the pointer depending on the eigenvalue of the observable $\sigma$. Let us denote the initial state of the pointer by its wavefunction $\phi(q)$. Then the action of the unitary on the eigenstates $\ket{\uparrow}$ and $\ket{\downarrow}$ of $\sigma$ is simply
\begin{align}
	e^{i \sigma \otimes p} \ket{\uparrow} \otimes \ket{\phi(q)} &= \ket{\uparrow} \otimes \ket{\phi(q-1)}, \\
	e^{i \sigma \otimes p} \ket{\downarrow} \otimes \ket{\phi(q)} &= \ket{\downarrow} \otimes \ket{\phi(q+1)}.
\end{align}

More generally, if the spin-$\half$ particle begins in a state described by the density matrix $\rho$, the post-measurement state of the system and pointer is found to be
\begin{align}\label{finalstate}
	\hat{U} \big( \rho \otimes \ket{\phi(q)}\!\bra{\phi(q)} \big) \hat{U}^\dagger = \; &\pi^+\!\rho\pi^+ \otimes \ket{\phi(q-1)}\!\bra{\phi(q-1)} + \pi^-\!\rho\pi^- \otimes \ket{\phi(q+1)}\!\bra{\phi(q+1)} \nonumber\\
	&\,+ \pi^+\!\rho\pi^- \otimes \ket{\phi(q-1)}\!\bra{\phi(q+1)} + \pi^-\!\rho\pi^+ \otimes \ket{\phi(q+1)}\!\bra{\phi(q-1)},
\end{align}
where $\pi^+=\ket{\uparrow}\!\bra{\uparrow}$ and $\pi^-=\ket{\downarrow}\!\bra{\downarrow}$, the projectors onto the eigenstates of $\sigma$, and $\ket{\phi(q\pm 1)}$ is the original pointer state displaced by the eigenvalue $\pm 1$.

To quantify the disturbance to the system, we calculate its post-measurement state by tracing out the pointer. (Here we trace over all pointer positions, ignoring the outcome of the measurement. For the outcome dependent post-measurement state, see Appendix D.)
\begin{equation}\label{firstpost}
	\rho^\prime = \pi^+\!\rho\pi^+ + \pi^-\!\rho\pi^- + \pi^+\!\rho\pi^- \braket{\phi(q+1)|\phi(q-1)} + \pi^-\!\rho\pi^+ \braket{\phi(q-1)|\phi(q+1)}.
\end{equation}

The effect of the pointer's initial state on the system's post-measurement state is only via the scalar product $\braket{\phi(q+1)|\phi(q-1)}$, which we denote as
\begin{equation}\label{qualfact}
	F[\phi] e^{i\Theta[\phi]} = \braket{\phi(q+1)|\phi(q-1)} = \int_{-\infty}^{+\infty} \!\!\!\!\! dq \;\;\; \phi^*(q+1) \;\phi(q-1).
\end{equation}

Here $F$ and $\Theta$ are both real valued functionals of the pointer's initial state. $F[\phi]$ is between $0$ and $1$, with $F=0$ corresponding to a strong measurement, in which the system is completely decohered in the eigenbasis of the observable, and $F=1$, ($\Theta=0$) corresponding to the state not being disturbed at all.

The phase $\Theta[\phi]$ does not reflect a disturbance of the state, rather only a change in the relative phase of the eigenstates of the observable. To see this, we construct a unitary operator $R$, defined by its effect on the eigenstates, $R\ket{\uparrow} = \ket{\uparrow}$ and $R\ket{\downarrow} = e^{i\Theta[\phi]} \ket{\downarrow}$. Equivalently, $R = \pi^+ + e^{i\Theta[\phi]} \pi^-$, and the post-measurement state of the system in Eq. (\ref{firstpost}) can be expressed as
\begin{equation}\label{poststate}
	\rho^\prime = R \Big[ F[\phi] \cdot  \rho  + (1-F[\phi]) \cdot \left( \pi^+\!\rho\pi^+ + \pi^-\!\rho\pi^- \right) \Big] R^\dagger.
\end{equation}

Thus the effect of the measurement is in two stages, a partial decoherence in the eigenbasis of the observable, followed by a unitary operation. As the proportion of the original state that is not decohered is $F[\phi]$, we label it the `quality factor' of the measurement.

In the case of a real pointer, the phase $\Theta[\phi]=0$, and $R$ is the identity operator. In this case we recover Eqs. (4) and (5) of the main text,
\begin{align}\label{realpointer}
	\rho^\prime &= F[\phi] \cdot  \rho  + (1-F[\phi]) \cdot \left( \pi^+\!\rho\pi^+ + \pi^-\!\rho\pi^- \right), & F[\phi] &= \int_{-\infty}^{+\infty} \!\!\!\!\! dq \;\;\; \phi(q+1) \;\phi(q-1).
\end{align}

Next, we determine the probabilities of the outcomes of the measurement.  A measurement on a spin-$\half$ particle can have only two possible outcomes. A natural way to map these outcomes onto a continuous pointer is to associate all positive positions to the outcome $+1$, and all negative positions to the outcome $-1$ (see discussion in Part C). Tracing over the system and positive pointer positions in Eq. (\ref{finalstate}), we obtain the probability of the $+1$ outcome
\begin{equation}
	P(+) = P(q>0) = tr( \pi^+ \!\rho ) \int_{0}^{+\infty} \!\!\!\!\! dq \; \left|\phi(q-1)\right|^2 + tr( \pi^- \!\rho ) \int_{0}^{+\infty} \!\!\!\!\! dq \; \left|\phi(q+1)\right|^2.
\end{equation}

In this work, we consider pointer states of symmetric modulus, i.e. $|\phi(q)|=|\phi(-q)|$. In this case we have that $\int_{0}^{\infty} dq \; \left|\phi(q)\right|^2 = \frac{1}{2}$, and $\int_{0}^{1} dq \; \left|\phi(q)\right|^2 = \int_{-1}^{0} dq \; \left|\phi(q)\right|^2 = \frac{1}{2} \left( \int_{-1}^{1}  dq \; \left|\phi(q)\right|^2 \right)$. With these simplifications, the probabilities for both of the outcomes take the form
\begin{equation}\label{prob2}
	P(\pm) = \frac{1}{2} \left[ 1\pm tr(\sigma\rho) \int_{-1}^{+1} \!\! dq \; \left|\phi(q)\right|^2 \right] = \frac{1}{2} \Big[ 1 \pm G[\phi] tr(\sigma\rho) \Big],
\end{equation}
where $\sigma=\pi^+-\pi^-$ is the spin observable, and we have labelled the integral as $G[\phi]$, another functional of the pointer's initial state,
\begin{equation}\label{Gphi}
	G[\phi] = \int_{-1}^{+1} \Mod{\phi(q)}^2 dq
\end{equation}

To understand the role of $G[\phi]$ we re-express the probabilities of the outcomes, and recover Eq. (6) of the main text,
\begin{equation}\label{precision}
	P(\pm) = G[\phi] \cdot \frac{1}{2} \Big[ 1 \pm tr(\sigma\rho) \Big] + (1-G[\phi]) \cdot \frac{1}{2}.
\end{equation}

The first term is the probability expected from a strong quantum measurement, multiplied by $G[\phi]$, while the second term represents a random guess (probability of both outcomes equal to $1/2$), multiplied by $1-G[\phi]$. This motivates the labelling of $G[\phi]$ as the `precision' of the weak measurement.

\newpage
\section{B. Determining the optimal pointer.}\label{AppB}

Here we use variational calculus to determine the highest precision $G[\phi]$ (Eq. \ref{Gphi}) for a given quality factor $F[\phi]$ (Eq. \ref{qualfact}) of a pointer state. As stated, we work with states of symmetric modulus, $\Mod{\phi(q)} = \Mod{\phi(-q)}$. Using the method of Lagrange multipliers to capture the constraints of: (i) the normalization of the pointer, and (ii) the quality factor $F$, we obtain the real-valued functional
\begin{align}
	\mathbf{S} = \int_{-1}^{+1} \!\!\!\!\! dq \; \left|\phi(q)\right|^2 &+ \lambda_1 \left[ \int_{-\infty}^{+\infty} \!\!\!\!\! dq \; \left|\phi(q)\right|^2 - 1 \right] \nonumber\\
	&+ \lambda_2 \left[ \left( \int_{-\infty}^{+\infty} \!\!\!\!\! dq \;\;\; \phi^*(q+1) \; \phi(q-1) \right) \left( \int_{-\infty}^{+\infty} \!\!\!\!\! dq^\prime \;\;\; \phi(q^\prime+1) \; \phi^*(q^\prime-1) \right) - F^2 \right].
\end{align}

The modulus $F^2$ is used rather than $Fe^{i\Theta}$ to ensure that the functional $\mathbf{S}$ is real. To find the extremum of $\mathbf{S}$ w.r.t. $\phi$, $\lambda_1$ and $\lambda_2$, consider that $\phi(q) = \phi_0(q) + \epsilon\eta(q)$, where $\phi_0(q)$ is the required optimal pointer wavefunction, $\epsilon$ is a small complex number, and $\eta(q)$ is another function that obeys $\eta(-\infty)=\eta(+\infty)=0$. The condition that $\phi_0(q)$ maximizes $\mathbf{S}$ implies that
\begin{equation}
	\frac{\partial \mathbf{S}}{\partial \epsilon}\bigg|_{\epsilon=0} = \frac{\partial \mathbf{S}}{\partial \epsilon^*}\bigg|_{\epsilon*=0} = 0 \;\;\; \forall \;\eta(q).
\end{equation}

This condition yields the following piecewise equation that determines the optimal pointer wavefunction,
\begin{equation}\label{opteq}
	\begin{cases}
	2 \left[ 1 + \lambda_1 \right] \phi_0(q) + \lambda_2 F \left[ e^{-i\Theta} \phi_0(q-2) + e^{i\Theta} \phi_0(q+2) \right] = 0 & \text{if } -1<q<+1, \\
	\;\;\;\;\;\;\;\;\;\;\; \lambda_1 \phi_0(q) + \lambda_2 F \left[ e^{-i\Theta} \phi_0(q-2) + e^{i\Theta} \phi_0(q+2) \right] = 0 & \text{otherwise.}
	\end{cases}
\end{equation}

Since the value of $\phi_0(q+2)$ is determined solely by $\phi_0(q)$ and $\phi_0(q-2)$, the entire wavefunction can be deduced from its value in any interval of length $4$ units. To begin with, consider that the wavefunction $\phi_0(q)$ is known in the interval $-1<q\leq3$. Then the wavefunction in the interval $q\in (3,\infty)$ can be contructed from the following sequence of complex functions $f_n:(-1,1]\rightarrow\mathbb{C}$, defined through the iterative application of Eq. (\ref{opteq})

\begin{align}\label{waveseq}
	f_n(x) &=	\begin{cases}
				\phi_0(x) \!\!\!& \text{if} \; n=0 \\
				\phi_0(x+2) \!\!\!& \text{if} \; n=1 \\
				-\frac{\lambda_1}{\lambda_2 F} e^{-i\Theta} f_{n-1}(x) - e^{-2i\Theta} f_{n-2}(x) \!\!\!& \text{if} \; n \geq 2 
				\end{cases}
\end{align}

Thus for $2n\minus 1\!<\!q\!\leq\!2n\!+\!1$, the wavefunction $\phi(q) = f_n(q-2n)$. Note that we have excluded the case of $\lambda_2 F=0$. This special case has only one normalizable solution, that $\phi_0(q)$ is a wavefunction entirely contained within the interval $[-1,1]$. This corresponds to the quality factor $F=0$, and the precision $G=1$, which is a measurement of maximum disturbance, and maximum information gain.

To check that the wavefunction in Eq. (\ref{waveseq}) is normalizable, we calculate its norm over the interval $(-1,\infty)$.
\begin{align}
	\int_{-1}^\infty \left| \phi(q) \right|^2 dq &= \int_{-1}^{1} + \int_{1}^{3} + ... \left| \phi(q) \right|^2 dq = \sum_{n=0}^\infty \int_{-1}^1 \left| f_n(x) \right|^2 dx
\end{align}

For this sum to converge it must be that $\lim_{n\rightarrow\infty} f_n(x) = 0$. As we will show next, this implies that the sequence $\Mod{f_n(x)}$ must be described by a decreasing exponential.

\line(1,0){455}
\bigskip
\twocolumngrid

\emph{Lemma. ---} For a sequence of complex numbers $s_n$, where $n\geq0$, that obeys the relation
\begin{equation}
	e^{-i\Theta} s_n + e^{i\Theta} s_{n+2} = \gamma s_{n+1} \;\;\;\;\;\;\; \gamma \in \mathbb{C}, \Theta \in \mathbb{R},
\end{equation}
$\lim_{n\rightarrow\infty} s_n =0$ if and only if $s_n \propto e^{-a n}e^{-in\Theta}$, where $a$ is defined by $\gamma = e^a + e^{-a}$, and Re$(a)>0$.

\emph{Proof. ---} Clearly if $s_n\propto e^{-a n}e^{in\Theta}$ with Re$(a)>0$, then $\lim_{n\rightarrow\infty} s_n =0$, proving sufficiency. To prove necessity, rewrite $\gamma = e^a + e^{-a} = 2\cosh a$, where $a\in\mathbb{C}$. This is always possible because the range of $\cosh$ is the entire complex plane. In fact, from the symmetry of $\cosh$, we can assume $\text{Re}(a)\geq 0$.

For $s_n$ to converge to zero, any linear combination of terms of the form $b_1 s_n + b_2 s_{n+m}$ must also converge to zero for fixed $b_1,b_2\in\mathbb{C}$, and $m\in\mathbb{N}$. In particular, consider the expression
\begin{align}
	\Delta_n &= e^{i\Theta} s_{n+2} - e^{-i\Theta} s_n.
\end{align}

By induction, it can be shown that $\Delta_n$ is given by
\begin{align}
	\Delta_n &= e^{n(a-i\Theta)} \left( s_1 e^a - s_0 e^{-i\Theta} \right) \nonumber\\
	&\;\;\;\;\;\;\; + e^{-n(a+i\Theta)} \left( s_1 e^{-a} - s_0 e^{-i\Theta} \right). \label{seqdiff}
\end{align}

(The form for general $n$ can be proved by induction). If $\text{Re}(a)>0$, $\Delta_n$ can only converge to zero if the coefficient multiplying the divergent term $e^{na}$ is zero. Thus $s_1 = s_0 e^{-a-i\Theta}$. In this case one can show that the entire sequence is described by $s_n = s_0 e^{-n(a+i\Theta)}$, as claimed in the lemma.

If $\text{Re}(a) = 0$, then $\Delta_n$ cannot converge. To see this, express $a = i\omega$ for some $\omega\in\mathbb{R}$, and
\begin{align}
	\left| \Delta_n \right| &= \Big| e^{in\omega} \left( s_1 e^{i\omega} - s_0 e^{-i\Theta} \right) \nonumber\\
	&\;\;\;\;\;\;\; + e^{-in\omega} \left( s_1 e^{-i\omega} - s_0 e^{-i\Theta} \right) \Big|.
\end{align}

If $\omega$ is a rational multiple of $\pi$, i.e. $\omega/\pi \in \mathbb{Q}$, one can find an infinite unbounded sequence of integers $n$ s.t. $n\omega$ is a multiple of $2\pi$, for which $|\Delta_n| = |\Delta_0|$. Thus the convergence of $\Delta_n$ implies that $\Delta_0=0$, which results in $s_1 = s_0 e^{-i\Theta} \cos \omega$. In this case, the sequence can be explicitly calculated to be $s_n = s_0 e^{-in\Theta} \cos n\omega$ which is not convergent for any $\omega\in\mathbb{R}$. If $\omega$ is an irrational multiple of $\pi$, then using Hurwitz's theorem, one can find an infinite unbounded sequence of integers $n$ s.t. $n\omega$ is arbitrarily close to a multiple of $2\pi$. This induces the same condition for convergence, $\Delta_0=0$, which leads to the above mentioned non-convergent sequence.

Applying this lemma to the sequence of functions $f_n(x)$ in Eq. (\ref{waveseq}) that we use to construct the optimal pointer state, we find that in order that the state is normalizable, it must be that
\begin{align}
	f_n(x) &= e^{-na} e^{-in\Theta} \phi_0(x) \;\;\;\;\; x\in(-1,1], \;\;\; n\geq 0, \\
	\text{where} \; &e^a + e^{-a} = -\frac{\lambda_1}{\lambda_2 F}, \;\;\; \text{Re}(a)>0.
\end{align}

Repeating this procedure for the interval $q\in(-\infty,+1)$, one finds a similar exponential envelope is necessary for the state to be normalizable. Thus the optimal wavefunction is described by an arbitrary normalizable wavefunction $f(x)$ in the interval $(-1,1]$ (whose modulus $|f(x)|$ is symmetric, since the functionals $F$ and $G$ have been derived under this assumption), copied piecewise to every other interval and modulated by the appropriate exponential factor
\begin{align}\label{newform}
	\phi_0(q) &= \begin{cases}
		f(q) & \forall q\in(-1,+1) \\
		f(q\!-\!2n) e^{-|n|a} e^{-in\Theta} &\forall q\in(2n\!-\!1,2n\!+\!1), \nonumber
	\end{cases} \\
	&\text{where} \;\;\;\;\; e^a + e^{-a} = -\frac{\lambda_1}{\lambda_2 F}, \;\;\;\;\; \text{Re}(a)>0.
\end{align}

At this stage one can conclude that the wavefunction must be equal to zero at all points of the form $q=2n-1$. Consider the wavefunction in the neighbourhood of $q=1$. From Eq. (\ref{newform}),
\begin{align}
	\lim_{q\rightarrow 1^+} \left| \phi_0(q) \right| &= e^{\text{Re}(a)} \lim_{q\rightarrow -1^+} \left| \phi_0(q) \right|, \\
	\lim_{q\rightarrow -1^+} \left| \phi_0(q) \right| &= \lim_{q\rightarrow 1^-} \left| \phi_0(q) \right| \;\;\; \text{from symmetry,} \\
	\therefore \lim_{q\rightarrow 1^+} \left| \phi_0(q) \right| &= e^{\text{Re}(a)} \lim_{q\rightarrow 1^-} \left| \phi_0(q). \right|
\end{align} 

Thus the left and right side limits of $\phi_0(q)$ at $q=1$ are unequal unless they are both equal to zero. Since a wavefunction must be continuous, the left and right side limits must be equal, and thus must be zero. This argument is then repeated for all other points of the form $q = 2n-1$.

To determine the precision $G[\phi_0]$ and quality factor $F[\phi_0]$ for this state, we use the form in Eq. (\ref{newform}). Recalling that the precision $G$ (Eq. \ref{Gphi}) is just the norm of the wavefunction in the interval $(-1,1)$,
\begin{equation}
	G[\phi_0] = \int_{-1}^1 \left| \phi_0(q) \right|^2 dq = \int_{-1}^1 \left| f(q) \right|^2 dq,
\end{equation}
one finds for the entire norm,
\begin{align}
	\int_{-\infty}^\infty \left| \phi_0(q) \right|^2 dq &= G[\phi_0] \frac{1+e^{-2\text{Re}(a)}}{1-e^{-2\text{Re}(a)}} \\
	\therefore G[\phi_0] &= \frac{1-e^{-2\text{Re}(a)}}{1+e^{-2\text{Re}(a)}}. \label{optG}
\end{align}

Repeating the same for the quality factor and phase $Fe^{i\Theta}$,
\begin{align}
	F[\phi_0] e^{i\Theta} &= \int_{\infty}^\infty \phi_0^*(q+1)\phi_0(q-1) dq \\
	&= e^{i\Theta} \frac{2e^{-\text{Re}(a)}\cos \text{Im}(a)}{1-e^{-2\text{Re}(a)}}.
\end{align}

Since $G[\phi_0]$ is only dependent on Re$(a)$, we are free to choose Im$(a)$ to maximize the quality factor $F[\phi_0]$. Choosing Im$(a)=0$ suffices to maximize $F[\phi_0]$,
\begin{equation}
	F[\phi_0] = \frac{2 e^{-\text{Re}(a)}}{1+e^{-2\text{Re}(a)}}. \label{optF}
\end{equation}

One can now calculate the trade-off between the precision (\ref{optG}) and quality factor (\ref{optF}) for an optimal pointer,
\begin{equation}\label{opttradeoff}
	F[\phi_0]^2 + G[\phi_0]^2 = 1.
\end{equation}

Importantly, since the function $f(q)$ used to generate the wavefunction of the optimal pointer in Eq. (\ref{waveseq}) in the central interval $(-1,1)$ is arbitrary (upto to symmetric modulus), there exists an entire family of optimal pointers for every chosen precision $G$. Every member of this family achieves the optimal trade-off $F^2+G^2=1$. Since we have the freedom of choosing the function $f(q)$ to be real and the phase $\Theta=0$ while maintaining the same $\{F,G\}$, we conclude that complex pointers do not provide any advantage as regards the trade-off.

\begin{figure}[h]
\includegraphics[width=0.8\linewidth]{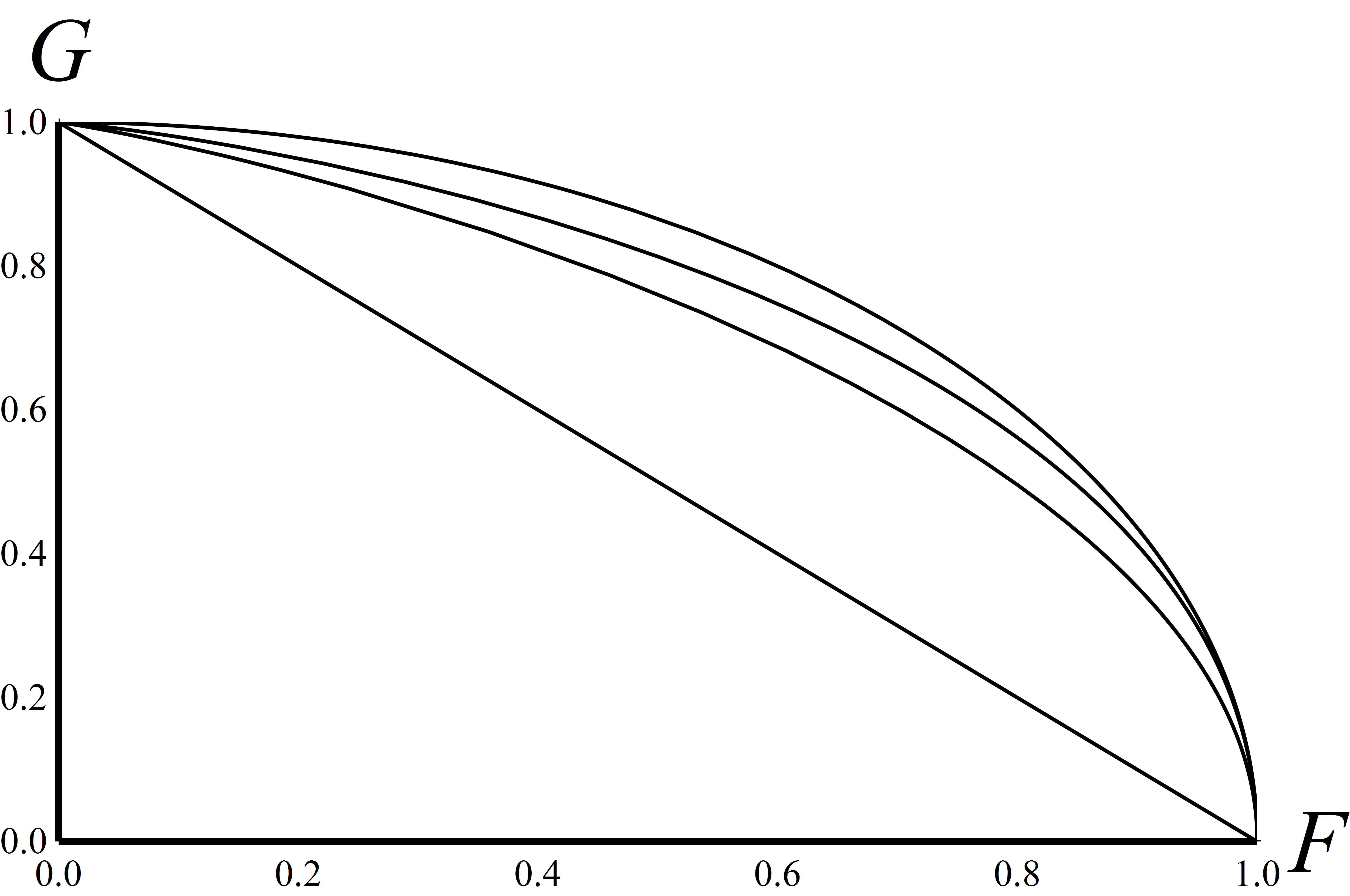}
\caption{\label{comparison} The trade-off between the precision $G$ and quality factor $F$ for various pointers: (from bottom to top) square, Gaussian, exponential, and the optimal pointer.}
\end{figure}

A comparison of the trade-off provided by this family of optimal pointers versus that achieved by various other pointer types is provided in Fig \ref{comparison}.

\emph{Additional constraints. ---} One may wish to apply further constraints on the form of the optimal wavefunction, such as continuity or continuous differentiability, and ask whether it is still possible to achieve any desired values of $\{F,G,\Theta\}$ under such constraints.

In fact, there exist families of optimal pointers that are infinitely differentiable, for instance, by choosing $f(q)$ in the central interval $(-1,1]$ to be
\begin{equation}
	f(q) = A e^{-\alpha/(1-q^2)}, \;\;\; q\in(-1,1], \;\;\; \alpha \in \mathbb{R}
\end{equation}

This function has the property of its one-sided derivatives being zero at the points $q=-1^+$ and $q=+1^-$, and thus the optimal pointer generated by copying this function to all other intervals will be infinitely differentiable, and able to achieve any optimal pair of $\{F,G\}$ by tuning the parameters $\{A,\alpha\}$.

\emph{The worst pointer.---} For completeness, we ask what is the worst possible pointer, i.e. one that maximally disturbs the state for any given precision of the measurement.  Clearly, the quality factor $F$ must be zero, implying that the overlap $\int\phi^*(q+1)\phi(q-1)\,dq=0$. A simple manner of constructing such a state is to take any member of the family of optimal pointers defined in Eq. (\ref{newform}), set $\phi(q)$ in every alternate region to be zero (i.e. in the regions $-3\!<\!q\!<\!-1$, $1\!<\!q\!<\!3$, $5\!<\!q\!<\!7$, etc.), and re-normalize the state.

\onecolumngrid
\newpage
\section{C. Reading the pointer state in a dichotomic von Neumann measurement.}\label{AppC}
\twocolumngrid

The nature of the von-Neumann coupling in a measurement on a spin-$\half$ particle is that when the spin is $+\half$ the pointer shifts by a positive value, and when the spin is $-\half$, it shifts by the same negative value. It is therefore a natural choice that one reads the sign of the position of the pointer, and associates positive values with positive spin, and negative values with negative spin. However, this is not in general the best way of extracting the information stored in the pointer state, as we could associate the pointer position to the outcomes $\pm 1$ in a different manner, or even choose to read the outcome via a measurement on the pointer of an observable different from the position. The quality factor F that we define is unaffected by the choice of how we read the pointer, while the precision G is itself defined by this choice.

In this section, we will demonstrate two important points regarding the reading of the pointer state. Firstly, we show that if we are restricted to measurements of the pointer position, the method of associating the outcome to the sign of the position is the optimal strategy for a natural class of initial pointer states, including the Gaussian and our class of optimal pointer states. Furthermore, in the case of the optimal pointer states, we demonstrate that even allowing any observable on the pointer state to be measured, the strategy of reading the sign of the position is indeed the best strategy, as it extracts the maximum possible information stored in the state of the pointer. This means that we do not lose out on the optimal information gain by restricting ourselves to this particular method of reading the pointer state.

After the system-pointer interaction, the pointer has two possible states: displaced to the right, $\phi_+(q)=\phi_0(q-1)$, or displaced to the left $\phi_-(q)=\phi_0(q+1)$. Assume first that we are restricted to measurements of the position $q$. In order to extract the most information out of the pointer state, our outcome $\pm 1$ must have as high a probability as possible of matching the corresponding state of the pointer. Thus if the pointer position is found to be $q$, the outcome is $+1$ if $\Mod{\phi_+(q)}^2>\Mod{\phi_-(q)}^2$, i.e. $\Mod{\phi_0(q-1)}^2>\Mod{\phi_0(q+1)}^2$, and analogously for the $-1$ outcome.

It is easily verified that for a Gaussian wave-packet centred about zero, as well as any other wave-function of symmetric modulus that is monotonically decreasing away from zero, it is true that  $\Mod{\phi_0(q-1)}>\Mod{\phi_0(q+1)}$ for all $q>0$, and the opposite for $q<0$. Thus for such pointer states, the strategy of matching the outcome with the sign of the pointer maximizes the information gain.

However, even with this strategy, such states cannot achieve the same trade-off as the class of optimal pointers, since this class does not include functions monotonically decreasing from $q=0$ (such functions do not satisfy Eq. \ref{newform}). Thus we conclude that all wave-functions of symmetric modulus that are monotonic away from zero, such as a Gaussian wave-packet, cannot achieve the optimal trade-off under a measurement of only the pointer position.

As stated before, in order to extract as much information as possible from the pointer, one must maximize the probability of distinguishing the two possible states of the pointer. In fact, the probability $P_d$ of correctly distinguishing any two states is upper bounded by a simple expression dependent only on their scalar product, that we apply here to the pointer states:
\begin{equation}\label{optdist}
	P_d \leq \frac{1}{2} \left( 1 + \sqrt{ 1 - \Mod{\braket{\phi_+|\phi_-}}^2} \right) = \frac{1}{2} \left( 1 + \sqrt{ 1 - F^2} \right),
\end{equation}
since we have labelled the scalar product between the pointer states as $Fe^{i\Theta}$ (Eq. \ref{qualfact}). It can also be calculated that for our strategy of associating the outcome of the measurement to the sign of the position of the pointer, the probability of correctly distinguishing the pointer states turns out to be
\begin{equation}
	P_d = \frac{1}{2} \left( 1 + G \right).
\end{equation}

But since the class of optimal pointer states satisfies the relation $G = \sqrt{1-F^2}$, we see that the probability of distinguishing the pointer states correctly saturates the upper bound, Eq. (\ref{optdist}), and therefore this strategy of reading the pointer extracts the entire information available. Thus although our precision G is defined specific to the strategy that we employ, the optimal value of G is model independent.

\onecolumngrid
\newpage
\section{D. Relating the trade-off of information gain and disturbance to a Bell scenario.}\label{AppD}
\twocolumngrid

\begin{figure}[h]
\includegraphics[width=0.8\linewidth]{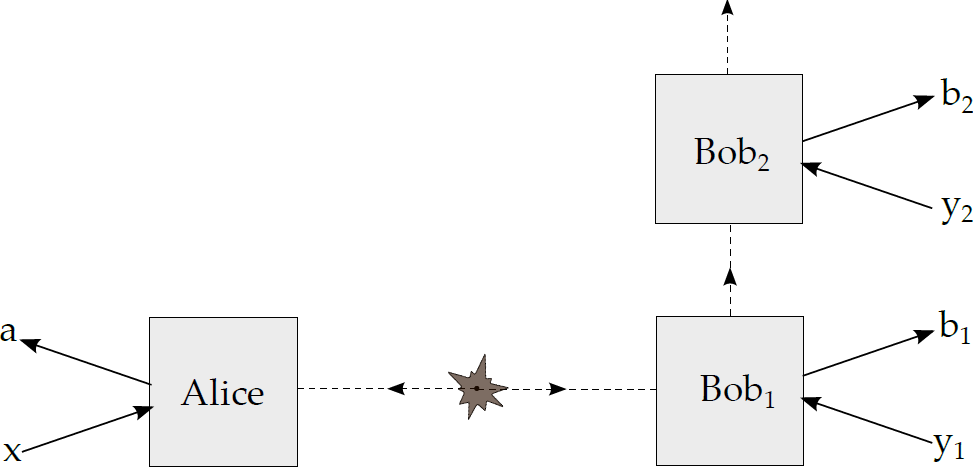}
\caption{\label{Bellfigure}Bell scenario involving a single Alice and multiple Bobs, where the dashed lines indicates a spin-$\half$ particle being transmitted, and the solid lines the inputs and outputs.}
\end{figure}

We study the Bell scenario depicted in Fig. \ref{Bellfigure}. Alice and Bob each possess one half of a singlet state of spin-$\half$ particles,
\begin{equation}
	\ket{\Psi} = \frac{\ket{\uparrow\downarrow} - \ket{\downarrow\uparrow}}{\sqrt{2}}.
\end{equation}

Alice receives a binary input $x\in\{0,1\}$, and accordingly performs a strong projective measurement of her spin along the corresponding direction $\bar{u}_x$; we label her outcome $a=\pm 1$. Bob receives two consecutive binary inputs $y_1,y_2\in\{0,1\}$, and performs two consecutive spin measurements along the corresponding directions $\bar{w}_{y_1}$ and $\bar{v}_{y_2}$ respectively; his outputs are labelled $b_1,b_2$ ($\pm 1$). Bob's first measurement has intermediate strength, while his second is a strong measurement.

To calculate the conditional probabilities $P(ab_1b_2|xy_1y_2)$, we simplify by imposing the no-signalling conditions between Alice and Bob, to get
\begin{equation}\label{Bellscenario2}
P(ab_1b_2|xy_1y_2) = P(a|x) P(b_1|xy_1a) P(b_2|xy_1y_2ab_1).
\end{equation}

For a projective measurement on one particle of a singlet pair, the probability of either outcome is $\half$. Thus $P(a|x) = \half$. After Alice's measurement, the state of the spin on Bob's side will be the projector on the spin in a direction opposite to Alice's post-measurement state,
\begin{equation}\label{singletsteer}
	\rho_{|xa} = \pi^{-a}_{\bar{u}_x} = \frac{1}{2} \left( I - a \bar{u}_x\cdit\bar{\sigma} \right),
\end{equation}
where $\pi^{\pm a}_{\bar{v}}$ represents the spin-$\half$ projector along the direction $\pm a\bar{v}$, and $\bar{\sigma}$ is the vector of Pauli matrices, $\{\sigma_x,\sigma_y,\sigma_z\}$.

The probability of the outcomes of Bob's first measurement, $P(b_1|xy_1a)$, is described by Eq. (\ref{precision}), a mixture of the probability of a strong quantum measurement and a random outcome. The probability in the case of a strong quantum measurement is $tr \left( \pi^{b_1}_{\bar{w}_{y_1}} \rho_{|xa} \right)$, and thus
\begin{equation}
	P(b_1|xy_1a) = \frac{1 - Gab_1\bar{x}\cdit\bar{y}_1}{2},
\end{equation}
where $G$ is the precision of Bob's first measurement. For the outcome of Bob's second measurement, we need to calculate the \emph{outcome dependent} post-measurement state of Bob's first measurement, $\rho_{|xy_1ab_1}$. This is done by tracing over the system and pointer in Eq. (\ref{finalstate}) for only positive or negative positions, corresponding to the $+1$ and $-1$ outcomes respectively.
\begin{align}
	\rho^\prime_{|xy_1ab_1} =\frac{F}{2}\rho_{|xa} &+ \left(\!\frac{1\!+\!b_1G\minus F}{2}\!\right)\pi^+_{\bar{w}_{y_1}}\!\rho_{|xa}\pi^+_{\bar{w}_{y_1}} \nonumber\\
	&+ \left(\!\frac{1\minus b_1G\minus F}{2}\!\right)\pi^-_{\bar{w}_{y_1}}\!\rho_{|xa}\pi^-_{\bar{w}_{y_1}}.
\end{align}

This state is not normalized, and its trace norm is precisely $P(b_1|xy_1a)$. From this state, we calculate $P(b_2|xy_1y_2ab_1) = tr(\pi^{b_1}_{\bar{v}_{y_2}} \rho^\prime_{|xy_1ab_1})$, and finally arrive at the complete expression for $P(ab_1b_2|xy_1y_2)$ as stated in Eq. (10) of the main text,
\begin{align}\label{condprob}
&P(ab_1b_2|xy_1y_2) = \frac{b_1G}{4} \left(\! \frac{a\; \bar{u}_x\!\cdit\bar{w}_{y_1} + b_2\; \bar{w}_{y_1}\!\cdit\bar{v}_{y_2}}{2} \!\right) +  \\
&\frac{F}{4} \left(\! \frac{1+ab_2\;\bar{u}_x\cdit\bar{v}_{y_2}}{2} \!\right) \!+\! \left(\!\frac{1\minus F}{4}\!\right) \!\left(\! \frac{1+ab_2\; \bar{u}_x\!\cdit\bar{w}_{y_1} \bar{w}_{y_1}\!\cdit\bar{v}_{y_2}}{2} \!\right) \nonumber
\end{align}

\onecolumngrid
\newpage
\section{E. Construction of the CHSH value of Alice and a sequence of Bobs.}\label{AppE}
\twocolumngrid

In this section, we will generalize the Bell scenario described in Fig. \ref{Bellfigure} to include a larger number of Bobs that have access, one after the other, to a particle from an entangled pair of spin-$\half$ particles, and demonstrate how to calculate the CHSH value of each of the Bobs with Alice.

Alice receives a binary input $x\in\{0,1\}$, and performs a strong measurement on her spin in the direction $\bar{u}_x$. Her outcome is labelled $a=\pm 1$. Each Bob receives an independent binary input, that we label $y_n\in\{0,1\}$ for Bob$_n$, and performs a spin measurement of intermediate strength on the particle in the corresponding direction $\bar{w}^{(n)}_{y_n}$. His outcome is labelled $b_n$. After measuring, the state is passed on to the next Bob in the sequence. Every Bob is ignorant of the input and the outcome of all of the preceding Bobs.

We would like to calculate the value of the CHSH expression for the statistics of Alice and Bob$_n$, defined as
\begin{equation}\label{CHSHexp}
	I^{(n)}_{CHSH} = E^{(n)}_{00} + E^{(n)}_{01} + E^{(n)}_{10} - E^{(n)}_{11}.
\end{equation}
Here $E^{(n)}_{xy_n}$ is the correlation of the outcomes of the measurements by Alice and Bob$_n$,
\begin{equation}\label{corrrho}
	E^{(n)}_{xy_n} = tr( \rho_n \; \sigma_{\bar{u}_x} \otimes \sigma_{\bar{w}_{y_n}}),
\end{equation}
where $\rho_n$ is the state of the pair of spin-$\half$ particles possessed by Alice and Bob$_n$, and $\sigma_{\bar{u}_x}$ and $\sigma_{\bar{w}_{y_n}}$ are the spin observables corresponding to the directions $\bar{u}_x$ and $\bar{w}_{y_n}$ respectively.

The correlation $E^{(n)}_{xy_n}$ can only depend on the following factors: the measurement settings of Alice and Bob$_n$, the state of the pair of spin-$\half$ particles that they share, and the precision $G_n$ of Bob$_n$'s measurement.

The contribution of the precision $G_n$ is straightforward. From Eq. (\ref{precision}), the probability for Bob$_n$'s outcome is a mixture of the probability as if the measurement was strong, along with a random outcome. The random contribution $(1-G)\cdot\half$ arises from the imprecision of the pointer's initial state, and is independent of the state on which the measurement is performed. Thus this term cannot be correlated in any way with the outcome of Alice. Thus the correlation only depends on the term corresponding to a strong measurement. More precisely,
\begin{equation}\label{corrG}
	E^{(n)}_{xy_n} = G_n \; \tilde{E}^{(n)}_{xy_n},
\end{equation}
where $\tilde{E}^{(n)}_{xy_n}$ denotes a correlation in the case of a strong measurement by Bob$_n$.

\onecolumngrid
\line(1,0){225}
\bigskip

Next, we require the state of the pair of spin-$\half$ particles before the measurements of Alice and Bob$_n$. Since this state depends on the inputs of all of the preceding Bobs, we label it $\rho_{n|y_1...y_{n-1}}$. To calculate it, we re-express the post-measurement state from Eq. (\ref{realpointer}) in the more compact form
\begin{equation}\label{Bellsplit}
	\rho_{n|y_1...y_{n-1}} = F_{n-1}\rho_{n|y_1...y_{n-2}} + (1-F_{n-1})D_{\bar{w}_{y_{n-1}}}(\rho_{n|y_1...y_{n-2}}),
\end{equation}
where $D_{\bar{v}}(\cdot)$ is a super-operator describing the decoherence of a state in the eigenbasis of the spin observable corresponding to the direction $\bar{v}$, as if the spin had been measured strongly along the direction $\bar{v}$,
\begin{equation}\label{decoherence}
	D_{\bar{v}}(\rho) = \pi^{+}_{\bar{v}} \rho\; \pi^{+}_{\bar{v}} + \pi^{-}_{\bar{v}} \rho\; \pi^{-}_{\bar{v}}.
\end{equation}
Here $\pi^{\pm }_{\bar{v}}$ represents a spin-$\half$ projector along the direction $\pm \bar{v}$. Thus each measurement splits the density matrix into a mixture of two density matrices, the first being the original undisturbed state, and the second corresponding to a decoherence operation by Bob$_n$ on the state, as if the state had been measured strongly by Bob$_n$. After $n-1$ Bobs have measured, the state will be a mixture of $2^{n-1}$ density matrices, each corresponding to decoherence operations by a different subset of Bobs. For example, the density matrix for Alice and Bob$_3$ has $4$ terms, corresponding to the two Bobs that have measured the state before Bob$_3$. If we denote the original entangled state as $\rho_1$, we have for Alice and Bob$_3$,
\begin{equation}\label{AB3}
	\rho_{3|{y_1y_2}} = F_1F_2\rho_1 + F_1(1-F_2)D_{\bar{w}_{y_2}}(\rho_1) + (1-F_1)F_2D_{\bar{w}_{y_1}}(\rho_1) + (1-F_1)(1-F_2)D_{\bar{w}_{y_2}}(D_{\bar{w}_{y_1}}(\rho_1)).
\end{equation}

More generally, in the state of Alice and Bob$_n$, there is only one term in the state (like the first term above) that corresponds to the density matrix of the original entangled state. The coefficient of this term is the product of the quality factors of all of the preceding Bobs' measurements, $\prod_{i=1}^{n-1} F_i$.

The rest of the terms each correspond to a different subset of the intermediate Bobs decohering the system via the super-operator $D_{\bar{v}}$. The coefficient of each such term will be a product of $F_i$ for every Bob$_i$ not in the subset, and $1-F_j$ for every Bob$_j$ in the subset. As an illustrative example, consider that in the density matrix of Alice and Bob$_7$, we pick the term corresponding to the subset $\{2,3,5\}$. This term appears in the state of Alice and Bob$_7$ as
\begin{equation}\label{example}
	F_1(1-F_2)(1-F_3)F_4(1-F_5)F_6 \; D_{\bar{w}_{y_5}}(D_{\bar{w}_{y_3}}(D_{\bar{w}_{y_2}}(\rho_1))).
\end{equation}

Now that we have the state of Alice and Bob$_n$ as a mixture of density matrices, we can find the correlation $E^{(n)}_{xy_n}$ by replacing each density matrix in the state by its corresponding correlation. For the single density matrix from the mixture that corresponds to the original entangled state, the correlation is as if Alice and Bob$_n$ measure on the original entangled state, and we denote this as $E^{(n)Q}_{xy_n}$. To simplify the correlation of the other terms, consider once again the case of Alice and Bob$_7$, and the density matrix corresponding to the subset $\{2,3,5\}$, which is $D_{\bar{w}_{y_5}}(D_{\bar{w}_{y_3}}(D_{\bar{w}_{y_2}}(\rho_1)))$. Using Eq. (\ref{corrrho}) and the definition of the super-operator in Eq. (\ref{decoherence}), we find the corresponding correlation to be
\begin{equation}
	E^{(2)Q}_{xy_2} \;\; (\bar{w}^{(2)}_{y_2}\cdot\bar{w}^{(3)}_{y_3}) (\bar{w}^{(3)}_{y_3}\cdot\bar{w}^{(5)}_{y_5}) (\bar{w}^{(5)}_{y_5}\cdot\bar{w}^{(7)}_{y_7}).
\end{equation}

We see that the correlation corresponding to a general subset of decohering Bobs is the product of the correlation of Alice and the first decohering Bob on the original state, followed by the scalar product of the measurement directions of each decohering Bob with the next (including the final Bob).

The state and correlation of Alice and Bob$_n$ that has calculated above are dependent on the inputs of all preceding Bobs. Since each Bob is ignorant of the input and outcome of the preceding Bobs, we have to sum over all possible choices of inputs with their corresponding probabilities. Thus the state of Alice and Bob$_n$ (without the dependence on preceding inputs) is
\begin{equation}\label{average}
	\rho_n = \sum_{y_1,...,y_{n-1}} \rho_{n|y_1...y_{n-1}} \;\prod_{i=1}^{n-1} P(y_n),
\end{equation}
where $P(y_n)$ is the probability of Bob$_n$ receiving the input $y_n$. The correlation $E^{(n)}_{xy_n}$ has to be averaged the same way.

\newpage
\section{F. A protocol for an arbitrary sequence of violations in the case of highly biased inputs.}\label{AppF}
\twocolumngrid

In this section we demonstrate that an arbitrary number N of Bobs can violate the CHSH inequality with Alice, in the case of unequal input bias, i.e. each Bob receives the one of the inputs $\{0,1\}$ more frequently than the other. We show this by constructing an explicit measurement protocol.

To begin with, we establish notation. As before, the measurement directions of Alice corresponding to the inputs $\{0,1\}$ will be denoted as $\{\bar{u}_0,\bar{u}_1\}$. The measurement directions of the $n^{th}$ Bob in the sequence, denoted Bob$_n$ are labelled $\{\bar{w}^{(n)}_0,\bar{w}^{(n)}_1\}$. The quality factor of Bob$_n$'s measurement is labelled as $F_n$. The precision $G_n$ of Bob$_n$'s measurement is calculated assuming he uses an optimal pointer (Eq. \ref{opttradeoff}), thus $G_n = \sqrt{1-F_n^2}$.

The initial entangled state of the two spin-$\half$ particles shared between Alice and the Bobs is taken to be the singlet. The protocol that they use has the following measurement directions,
\begin{align}\label{protocol}
	\mathbf{A}&\mathbf{LICE} & \mathbf{BO}&\mathbf{B_n} \\
	\bar{u}_0 &= -\bar{Z} & \bar{w}^{(n)}_0 &= \bar{Z} \nonumber \\
	\bar{u}_1 &= \bar{X} & \bar{w}^{(n)}_1 &= \cos\theta_n\bar{Z} + \sin\theta_n\bar{X} \nonumber,
\end{align} 
where $\bar{Z}$ and $\bar{X}$ are two orthogonal directions in space, and $\theta_n$ and the quality factor $F_n$ of Bob$_n$'s measurement are defined by the equations
\begin{align}
	\theta_1 &= \frac{\pi}{4}, \\
	\tan\theta_n &= \prod_{i=1}^{n-1} F_i, \\
	F_n &= 1 - \frac{2}{1+\sqrt{1+\tan^2\theta}} \label{choiceF},
\end{align}

Thus the measurement angle $\theta_n$ and the quality factor $F_n$ are both functions of the quality factors of all prior Bobs. From the initial angle $\theta=\pi/4$, it can be shown that both $\theta_n$ and $F_n$ are strictly decreasing sequences, and they satisfy
\begin{align}
	\lim_{n\rightarrow\infty} \theta_n &= \lim_{n\rightarrow\infty} F_n = 0 \\
	\text{but} \;\;\;\;\; \theta_n, \;F_n &>0 \;\;\; \forall n
\end{align}

Finally we account for the input bias of the Bobs in the sequence. Each Bob's input is independent of the other Bobs, and one can label by $r_k$ the probability that Bob$_k$ receives the input $1$. Thus $0<r_k<1$. In this calculation we only require the probability that at least one Bob prior to Bob$_n$ has received the input $1$, and we label this quantity $P_n$,
\begin{equation}
	P_n = 1 - \prod_{i=1}^{n-1} (1-r_k)
\end{equation}
Equivalently, $1-P_n$ is the probability that every Bob prior to Bob$_n$ has received the input $0$. Clearly, $P_n$ is a strictly increasing sequence, but always smaller than one. Importantly, $P_n$ can be made as small as one likes by simply making the individual biases $r_k$ small enough.

To calculate the CHSH value of Bob$_n$ with Alice in this protocol, we need to calculate the state after all of the prior Bobs have measured. The construction of the state and CHSH value for an arbitrary sequence of Bobs is detailed in Part E of this Supplemental Material. For our protocol, the state of Alice and Bob$_n$ is the weighted average of two cases, firstly that of all of the prior Bobs having received the input $0$, and the other corresponding to at least one Bob having received the input $1$.

In the first case, that of all prior Bobs receiving the input $0$, the state simplifies greatly because all of the Bobs share a common measurement setting corresponding to the input $0$, i.e. $\forall\;n,\;\bar{w}^{(n)}_{0} = \bar{Z}$. Consider the argument in section E (eqs \ref{Bellsplit}-\ref{example}), used to determine the state of Alice and Bob$_n$. The state is a mixture of $2^{n-1}$ density matrices, one of which is the original state, and the rest correspond to various subsets of Bobs having decohered the state, as if they had measured strongly. However, once a state is decohered in the $\bar{Z}$ direction, further decohering operations in the $\bar{Z}$ direction have no effect. Thus the decoherence operator $D_{\bar{Z}}$ from Eq. (\ref{decoherence}) obeys the relation $D_{\bar{Z}}\circ D_{\bar{Z}} = D_{\bar{Z}}$. Thus, in the case of all prior Bobs receiving the input $0$, the state of Alice and Bob$_n$ simplifies to
\begin{align}\label{limitrho}
	\rho^{(0)}_n &= \left( \prod_{i=1}^{n-1} F_i \right) \rho_1 + \left( 1 - \prod_{i=1}^{n-1} F_i \right) D_{\bar{Z}}(\rho_1) \\
	&= (\tan\theta_n) \rho_1 + (1-\tan\theta_n) D_{\bar{Z}}(\rho_1)
\end{align}
where $\rho_1$ is the original singlet state.

In the second case, that of at least one Bob having received the input $1$, the state splits into a similar expression as the above,
\begin{equation}\label{spoiltrho}
	\rho^{(1)}_n = \left( \prod_{i=1}^{n-1} F_i \right) \rho_1 + \left( 1 - \prod_{i=1}^{n-1} F_i \right) \rho^{\prime},
\end{equation}
where $\rho^{\prime\prime}$ is a density matrix corresponding to various Bobs having decohered the state, as if they had measured strongly. We do not calculate this state explicitly. However, since it involves at least one strong measurement by a prior Bob, it is a separable state between Alice and Bob$_n$.

Combining the two cases, weighted by their probabilities, one obtains the state of Alice and Bob$_n$,
\begin{align}
	\rho_n &= (1-P_n) \rho^{(0)}_n + P_n \rho^{(1)}_n \\
	&= (\tan\theta_n)\rho_1 + (1-P_n)(1-\tan\theta_n) D_{\bar{Z}}(\rho_1) \nonumber\\
	&\;\;\;\;\;\;\;\;\;\;\;\;\;\;\;\;\;\;\;\;\; + P_n(1-\tan\theta_n)\rho^\prime_n
\end{align}

One can now calculate the CHSH value of Alice and Bob$_n$ from the state, using Eqs. (\ref{CHSHexp}-\ref{corrrho}). First, calculating the CHSH values for only the singlet state $\rho_1$ and the decohered singlet $D_{\bar{Z}}(\rho_1)$, using the measurements of our protocol (Eq. \ref{protocol}),
\begin{align}
	\rho_1 \;:\;\;\; I^{(n)(1)}_{CHSH} &= 1 + \cos\theta_n+\sin\theta_n \\
	D_{\bar{Z}}(\rho_1) \;:\;\;\; I^{(n)(2)}_{CHSH} &= 1 + \cos\theta_n
\end{align}

For the state $\rho^\prime_n$, we lower bound its CHSH value by $-2$, the worst value for a separable state. Applying these CHSH values to the state of Alice and Bob$_n$, we obtain a lower bound for the final CHSH value,
\begin{align}
	I^{(n)}_{CHSH} &\geq G_n \big( \tan\theta_n (1+\cos\theta_n+\sin\theta_n) \nonumber\\
	&\;\;\;\;\;\;\;+ (1-P_n) (1-\tan\theta_n) (1+\cos\theta_n) \nonumber\\
	&\;\;\;\;\;\;\;+ P_n (1-\tan\theta_n) (-2) \big) \\
	&= G_n \left( 1 + \sec\theta_n - P_n (3+\cos\theta_n) (1-\tan\theta_n) \right) \\
	&\geq G_n \left( 1 + \sec\theta_n - 4 P_n \right),
\end{align}
where we have used $\theta_n\in(0,\pi/4]$ to bound the trigonometric coefficients of $P_n$.

Finally expressing $G_n=\sqrt{1-F_n^2}$ and replacing $1+\sec\theta_n$ from Eq. (\ref{choiceF}),
\begin{align}
	I^{(n)}_{CHSH} &\geq \sqrt{1-F_n^2} \left( \frac{2}{1-F_n} - 4P_n \right).
\end{align}

Thus a sufficient condition for $I^{(n)}_{CHSH}$ to be greater than the classical bound of $2$ is
\begin{equation}\label{condmult}
	P_n < \frac{1}{2} \left( \frac{1}{1-F_n} - \frac{1}{\sqrt{1-F_n^2}} \right).
\end{equation}

It is easily verified that the expression on the RHS, that we label $\chi_n$ is a strictly decreasing sequence, but greater than zero for all $n$. On the other hand, $P_n$ is a strictly increasing sequence. Thus, if there are $N$ Bobs and the input biases for each Bob is chosen such that for the final Bob$_N$, $P_N < \chi_N$, then it immediately follows that $P_{N-1}<P_N<\chi_N<\chi_{N-1}$, and by induction, $P_n<\chi_n$ for all $n\leq N$. Thus every Bob prior to Bob$_N$ will also violate the CHSH inequality with Alice.

It is always possible to pick the individual input biases $r_k$ to be small enough that $P_N<\chi_N$. However, $\chi_N$ is a decreasing sequence (and in fact can be shown to decrease quite rapidly), therefore the greater the number $N$ of Bobs that wish to violate the CHSH inequality with Alice, the smaller the probability $P_N$ must be, which in turn implies that the individual probabilities $r_k$ must be smaller. The unequal input bias is thus a critical feature of this protocol.

\onecolumngrid
\section{G. Behaviour of the CHSH violation for the protocol constructed in this work.}\label{AppG}
\twocolumngrid

We quantify the behaviour of the CHSH value for the measurement protocol constructed in Section F, applied to a long sequence of Bobs. For simplicity, we consider the limit that the probability that each Bob receives the input $1$, defined as $r_k$ for Bob$_k$, tends to $0$. In this limit, the state of Alice and Bob$_n$ is the state as if all the prior Bobs received the input $0$, which is state in Eq. (\ref{limitrho}). The CHSH value for this state using the measurement protocol of Eq. (\ref{protocol}) is simply
\begin{equation}
	I^{(n)}_{CHSH} = G_n (1 + \sec\theta_n ) = 2 \sqrt{\frac{1+F_n}{1-F_n}}.
\end{equation}

For large values of $n$, the quality factors $F_n<<1$, and we approximate the above to first order to obtain the CHSH value as
\begin{equation}
	I^{(n)}_{CHSH} \approx 2 (1+F_n)
\end{equation}

We can define the CHSH violation to be $V_n = I^{(n)}_{CHSH} - 2$. Thus for large $n$, $V_n \approx 2F_n<<1$. We proceed to calculate $F_n$ itself to first order from Eq. (\ref{choiceF}),
\begin{equation}
	F_n \approx \frac{\left( \prod_{i=1}^{n-1} F_i \right)^2}{4}.
\end{equation}

Combining these two results to calculate $F_{n+1}$, and then $V_{n+1}$, we find that, for large $n$,
\begin{align}
	V_{n+1} \approx 2F_{n+1} &\approx \frac{\left( \prod_{i=1}^{n} F_i \right)^2}{2} \\
	&= \frac{F_n^2 \left( \prod_{i=1}^{n-1} F_i \right)^2}{2} \\
	&\approx 2 F_n^3 \approx \frac{V_n^3}{4}
\end{align}

Since $V_n<<1$ for large $n$, the above relation describes a super-exponentially decreasing sequence.


\begin{thebibliography}{99}
\bibitem{footnote1} An arbitrary coupling $\int g(t) dt = g_0$ with an initial pointer state $\phi(q)$ is equivalent to using unit coupling $\int g(t) dt = 1$ and a scaled version of the pointer state $\phi^\prime(q) = \sqrt{g_0} \phi(g_0 q)$.
\bibitem{footnote2} Changing the frequency of inputs does not affect the measurement independence or the CHSH bound of any Bob. As each Bob is independent of the rest, it is as if he measures on the density matrix obtained by the weighted average over all possible inputs and outcomes of the prior Bobs.
\end{thebibliography}
\end{document}